**Title:**

Transition from Diffusive to Superdiffusive Transport in Carbon Nanotube Networks via Nematic Order Control

**Authors:**

Michael Wais*[,a,b], Filchito Renee G. Bagsican*[,c,d], Natsumi Komatsu[e,f], Weilu Gao[g], Kazunori Serita[c], Hironaru Murakami[c], Karsten Held[b], Iwao Kawayama[c,h], Junichiro Kono [e,i,j,a,c], Marco Battiato[a] and Masayoshi Tonouchi[c]

**Affiliations:**

a. Division of Physics and Applied Physics, School of Physical and Mathematical Sciences, Nanyang Technological University, Singapore
b. Institute for Solid State Physics, TU Wien, 1040 Vienna, Austria
c. Institute of Laser Engineering, Osaka University, Suita, Osaka 565-0871, Japan
d. *Current address*: Femtosecond Spectroscopy Unit, Okinawa Institute of Science and Technology Graduate University, Onna-son, Okinawa 904-0495, Japan
e. Department of Electrical and Computer Engineering, Rice University, Houston, Texas 77005, United States
f. *Current address*: Department of Chemical and Biomolecular Engineering, University of California, Berkeley, Berkeley, CA 94720, United States
g. Department of Electrical and Computer Engineering, University of Utah, Salt Lake City, Utah 84112, United States
h. *Current address*: Graduate School of Energy Science, Kyoto University, Kyoto 606-8501, Japan
i. Department of Physics and Astronomy, Rice University, Houston, Texas 77005, United States
j. Department of Material Science and NanoEngineering, Rice University, Houston, Texas 77005, United States

*These authors contributed equally to this work.

Corresponding authors: **tonouchi.masayoshi.ile@osaka-u.ac.jp, marco.battiato@ntu.edu.sg, filchitorenee.bagsican@oist.jp**

**Abstract**

The one-dimensional confinement of quasiparticles in individual carbon nanotubes (CNTs) leads to extremely anisotropic electronic and optical properties. In a macroscopic ensemble of randomly oriented CNTs, this anisotropy disappears together with other properties that make them attractive for certain device applications. The question however remains if not only anisotropy, but other types of behaviours are suppressed by disorder. Here, we compare the dynamics of quasiparticles under strong electric fields in aligned and random CNT networks using a combination of terahertz emission and photocurrent experiments and out-of-equilibrium numerical simulations. We find that the degree of alignment strongly influences the excited quasiparticles' dynamics, rerouting the thermalisation pathways. This is, in particular, evidenced in the high-energy, high-momentum electronic population (probed through the formation of low energy excitons via exciton impact ionization) and the transport regime evolving from diffusive to superdiffusive.

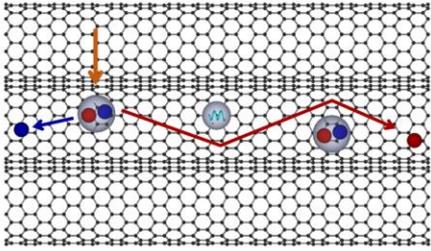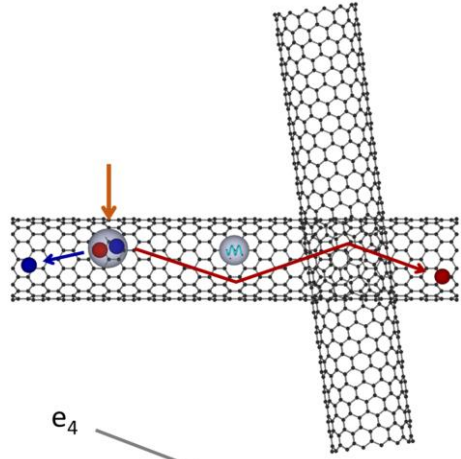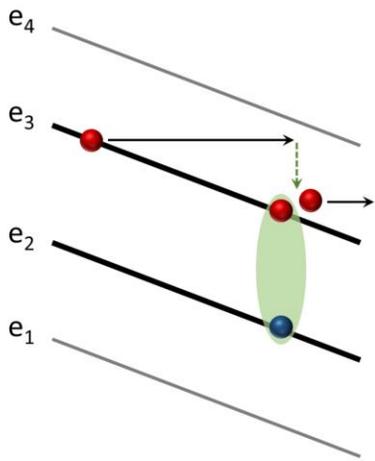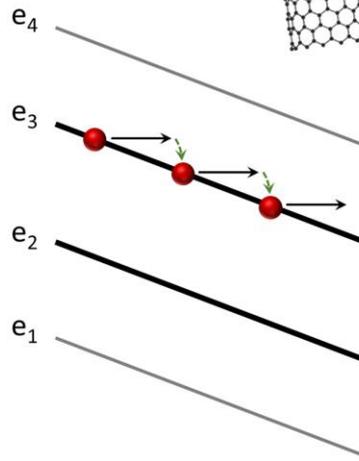

TOC Graphic

The emergence of low-dimensional materials has opened up exciting possibilities for exploring new physics and functionalities inherent to their dimensionality. Carbon nanotubes (CNTs) in particular, provide an ideal platform to study the behaviour of one-dimensional (1D) metals and semiconductors. Several groups have observed many complex condensed matter phenomena (Luttinger-liquid behaviour[1], Kondo effect[2], Aharanov-Bohm effect[3], superconductivity[4,5] and quantum conduction[6,7]) in CNTs and also demonstrated their potential for highly efficient devices[8,9]. Typically, the experiments are done with devices based on isolated CNTs or aligned rope-like structures (bundles) to allow access to dimension-related properties. In many applications, wide-area samples are preferable but the 1D properties could become inaccessible if the tubes are arranged in a random fashion. Macroscopic manifestations of these 1D properties have become recently achievable[10] due to advancements on assembly techniques that produce wafer-scale films of highly-aligned CNTs[11,12].

Despite such tremendous progress, much work is still needed to fully uncover the alignment mechanism and produce large samples of perfectly-aligned, crystalline CNTs of comparable structural quality to conventional bulk single crystals. As shown in the seminal work by He, *et. al*[11], the degree of alignment degrades when producing thicker films of metallic CNTs and some chiralities of CNTs also seem more difficult to align. Other alignment methods have also been proposed[13–17], with different advantages and disadvantages, and showing varying degrees of success. With all these difficulties, it seems logical to ask if achieving perfect alignment is really critical for optimal performance of CNT-based devices. But, even more importantly, if alignment can unlock qualitatively different dynamics that are instead suppressed in disordered CNT networks.

We attempt to answer these questions by comparing the dynamics of photo-generated quasiparticles under the influence of external electric field in aligned and random CNT networks. Specifically, we employ a combination of experiments (terahertz (THz) emission and photocurrent measurements) and numerical simulations to understand the relevant scattering processes in CNTs. The experimental techniques allow us to access the ultrafast current (within picosecond time scales) leading to THz radiation, and the slow current components through the total photocurrent. Combined with the numerical simulations that accurately model out-of-equilibrium processes[18], we have applied these techniques to unravel the roles of spontaneous exciton dissociation and exciton impact excitation in THz and photocurrent generation, respectively.

Furthermore, we show that electrons undergo superdiffusive (diffusive) motion in aligned (random) CNT networks, depending on the number of scatterings with phonons and impurities.

In the diffusive regime, momentum-randomising scatterings are frequent and the bias-induced acceleration only perturbatively alters the electronic population. Conversely, when momentum-randomising scatterings become rarer, the asymmetric displacement of the electronic population within the Brillouin zone becomes sizeable (leading to the so called superdiffusive regime) and can even trigger new thermalisation pathways that were previously forbidden due to energy or momentum thresholds. In the case of CNTs, we will show that this transport behaviour directly affects the kinetic energy gain of electrons by acceleration in the electric field, and the possibility of producing additional excitons by impact generation.

We fabricated dipole-type photoconductive antenna (PCA) structures (Fig. 1a) on top of the CNT films using standard photolithography techniques (see Supporting Information for details of device fabrication and experimental set-up). We prepared three CNT-based PCA devices that represent possible configurations: aligned CNT films with the tube axis direction parallel (Sample 1) and at 45° with respect to the electric field (Sample 2), and a randomly oriented film (Sample 3), as illustrated in Figs. 2a, 2b, and 2c, respectively. The CNT films had similar chirality purities (Fig. 1b) but with significantly different degrees of alignment. Specifically, based on the calculated values of the reduced linear dichroism ($LD^r$)[11], the aligned CNT samples had about twice as much better alignment compared with the random film (inset, Fig. 1b). The nematic order parameter ($S$) of Samples 1 and 2 was ~0.38; note that $S = 0$ for a completely random distribution and $S = 1$ for a perfectly aligned film[18,19].

The device configuration (*i.e.*, degree of alignment and direction relative to the external electric field) affects the carrier transport direction in the CNT network. Firstly, the effective bias voltage across each individual tube depends on its orientation with respect to the electric field (Figs. 1c and 1d, also see Supporting Information). Secondly, the electrical conductivity in the CNT film is highly anisotropic[20,21], with the ratio of conductivities between the parallel and perpendicular directions as high as sixty[11]. In a CNT network composed of interconnected nanotubes, the carriers will follow a path with the least electrical resistance, which is nominally determined by tube chirality, density, orientation, and connectivity[22–25]. In our devices, the red lines indicated in Figs. 2a-2c depict one probable path of carriers between the electrodes based on the degree of alignment and overall tube orientation relative to the electric field.

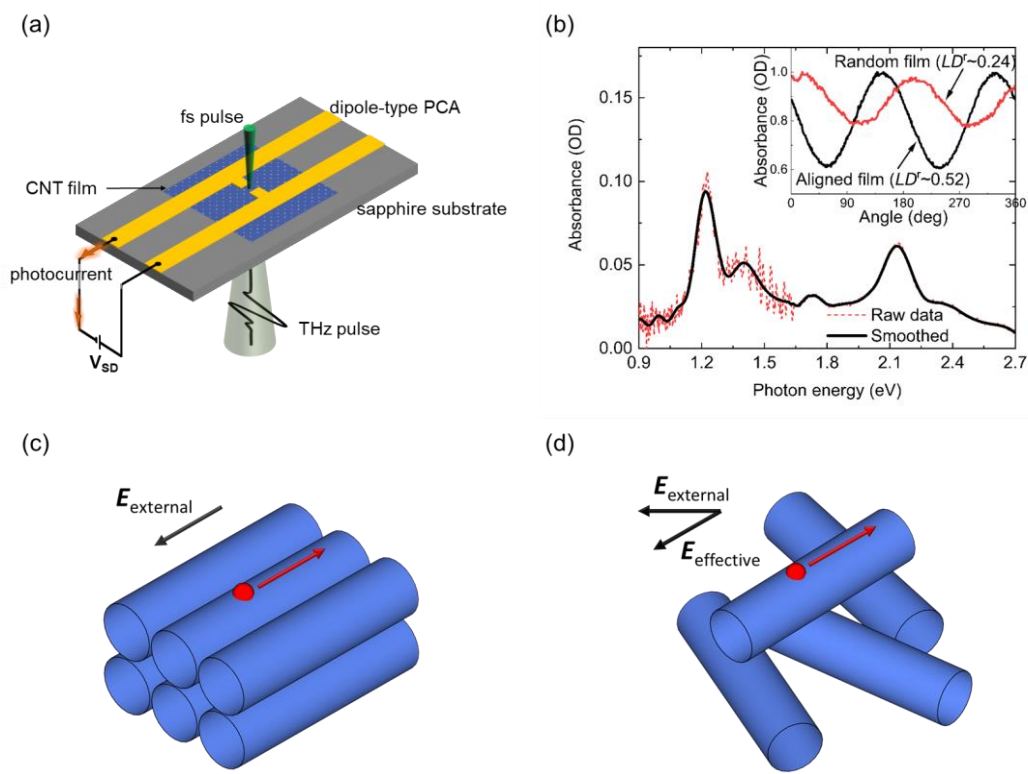

*Fig. 1.* (a) Device schematic diagram. (b) Unpolarized light absorption and polarization-dependent absorption at 660 nm (inset) of (6,5) CNT films. (c) Fully-aligned CNT film with electric field parallel to the alignment direction. (d) Completely random CNT film showing reduced effective electric field along the tube.

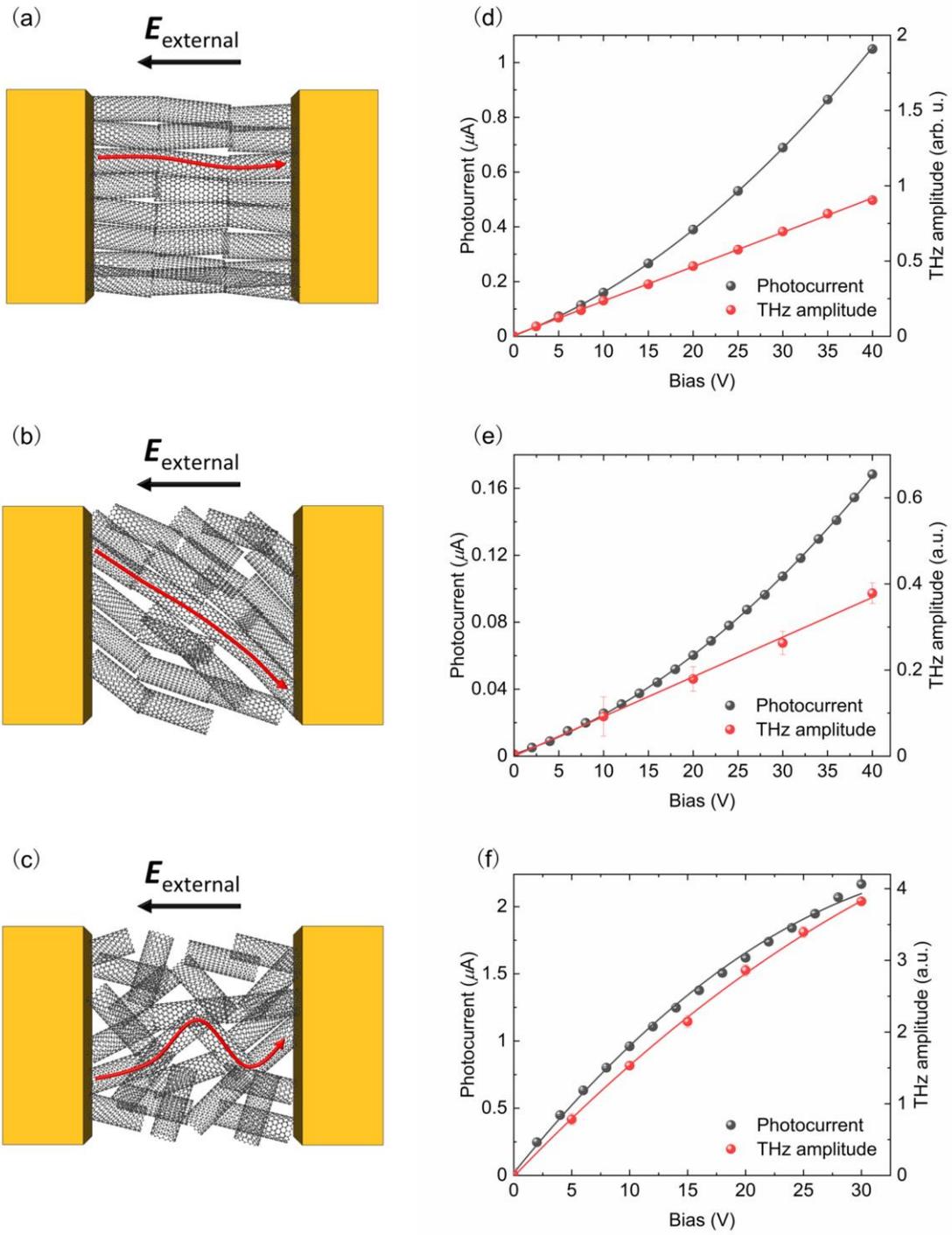

*Fig. 2.* Probable carrier conduction path and bias-dependence of THz amplitude and photocurrent for CNTs aligned parallel (a and d) and tilted relative to the electric field (b and e), and random CNTs (c and f), respectively.

Aside from the differences in the propagation path of carriers in the samples, the bias-dependence of THz emission and photocurrent have different behaviours as shown in Figs. 2d-2f. We observed a superlinear increase in photocurrent with bias and a linear increase in the THz amplitude with bias in both aligned samples (Samples 1 and 2) (Figs. 2d and 2e), whereas in Sample 3 (random CNT), both THz amplitude and photocurrent show a saturation-like sublinear behaviour with bias (Fig. 2f). To explain these differences, we employed a newly-developed numerical approach to the Boltzmann transport and scattering equation[18,26,27] to understand the mechanisms that contribute to THz and photocurrent generation in CNTs, the influence of the external electric field on the dynamics of photo-generated quasiparticles, and the effect of the degree of alignment and direction with respect to the electric field.

We considered two extreme cases in the current numerical simulations: a fully aligned CNT network with an electric field applied parallel to the alignment direction (Fig. 1c) and a completely random network of CNTs (Fig. 1d). Identical numerical parameters were implemented in the two cases, but we increased the electron–phonon (acoustic) scattering amplitude by ten times in the random CNT network to emulate a sharp increase in impurity-like scattering due to the multiple tube distortions at the tube intersections. Additionally, we account for the angle formed between the electric field and the CNTs in the random case (see Supporting Information for details).

The schematic diagram in Fig. 3a describes the processes occurring at different time scales for aligned and random CNTs, similar to what we have proposed previously[18]. In our simulations, we assume that the laser excitation only generates $E_{22}$ excitons and not free electron-hole pairs due to large difference in oscillator strengths[28] and also because our polarization- and excitation energy-dependence data indicate that both THz and photocurrent generation are initially triggered by $E_{22}$ exciton generation (*i.e.*, maximum responses occur for light linearly polarized parallel to the tube axis and at around 2.14 eV excitation, see Figs. S2 and S3 in the Supporting Information). The $E_{22}$ excitons then partially decay through phonon scatterings[29,30] into bright and dark $E_{11}$ excitons with similar energies but larger momenta.

These highly energetic excitons undergo further scatterings with phonons until they reach the bottom of their respective excitonic bands. Being charge-neutral, the excitons are not accelerated by the electric field and are more likely to survive longer (at least longer than the characteristic extraction time of 1 ps for carriers) before being annihilated through radiative or nonradiative recombination (this process is not included in the simulations). During this time, the high-energy tail of these excitons will slowly autoionize into additional free carriers due to the high temperature (Fig. 3a). The measured total photocurrent is the sum of this slow (*i.e.*, low-frequency) current (which does not lead to THz emission) and the ultrafast current that generates the THz emission.

The origin of this ultrafast current, on the other hand, is closely related to the initial $E_{22}$ excitons generated by laser excitation[18]. A fraction of these excitons autoionize into free electrons (holes) in the conduction (valence) bands through spontaneous dissociation[31], and their acceleration in the electric field and the subsequent extraction (turning-off) produce the impulsive current responsible for the THz emission ($\vec{E}_{\text{THz}} \propto \partial\vec{J}/\partial t$) (Figs. 4a and 4b). The same autoionization scattering amplitudes were used for both aligned and random CNTs. This means that the initial population of free carriers will be proportional to the number of absorbed photons and that the THz emission amplitude should linearly depend on the pump power, which we confirmed experimentally (see Fig. S4 in the Supporting Information).

One of the effects of randomness in the random CNT network is an increase in impurity-like scatterings arising from multiple tube distortions coming from the intersecting tubes (Fig. 1d). This leads to very different time evolution of the electronic band populations between aligned (Figs. 3b and 3c) and random (Figs. 3d and 3e) CNT networks. After the partial autoionization of $E_{22}$ excitons (*B* in Figs. 3b-3e), the free carriers are accelerated by the electric field, which leads to an asymmetry between positive and negative momenta in the electronic population. This acceleration is limited by momentum-randomising scatterings (electron-phonon and electron-defect)[32,33] that prevent indefinite gain of kinetic energy. While electron-phonon scatterings are expected to maintain approximately the same strength in both samples, the scatterings with lattice defects are amplified in random CNTs, leading to a diffusive-like (close to dynamic equilibrium) carrier transport (*C* and *D* in Figs. 3d and 3e), while we see a behaviour closer to ballistic carrier transport (and a very asymmetric distribution of electrons in $k$-space) for aligned CNTs (*C* and *D* in Figs. 3b and 3c).

The electron transport behaviour (diffusive or superdiffusive) within the first few picoseconds has important influence on the final number of excitons in the system, and to the photocurrent through the slow thermal autoionization of excitons surviving in the end (Fig. 3a). Exciton impact generation, wherein an electron loses energy by scattering with an impurity and generates a low-energy exciton in the process, becomes important for electrons having an energy above a certain threshold[18]. Due to energy conservation, this four-leg scattering channel (see Supporting Information for details) becomes active only for electrons with an energy larger than 1.2 eV above the band bottom. In random CNTs, the electrons' momentum is frequently randomised due to the increased impurity scatterings, preventing the carriers from reaching large momenta and therefore large energies, whereas the superdiffusive transport in aligned CNTs results in a significant number of electrons gaining sufficient energy to undergo exciton impact generation. The impact generation becomes favourable at high bias due to higher energy gain of electrons. Finally, the photocurrent in aligned CNTs is expected to show a superlinear dependence on bias (Fig. 5b) as a consequence of exciton impact generation (and the subsequent slow thermal autoionization) but a linear behaviour is expected in the case of random CNTs (Fig. 5a).

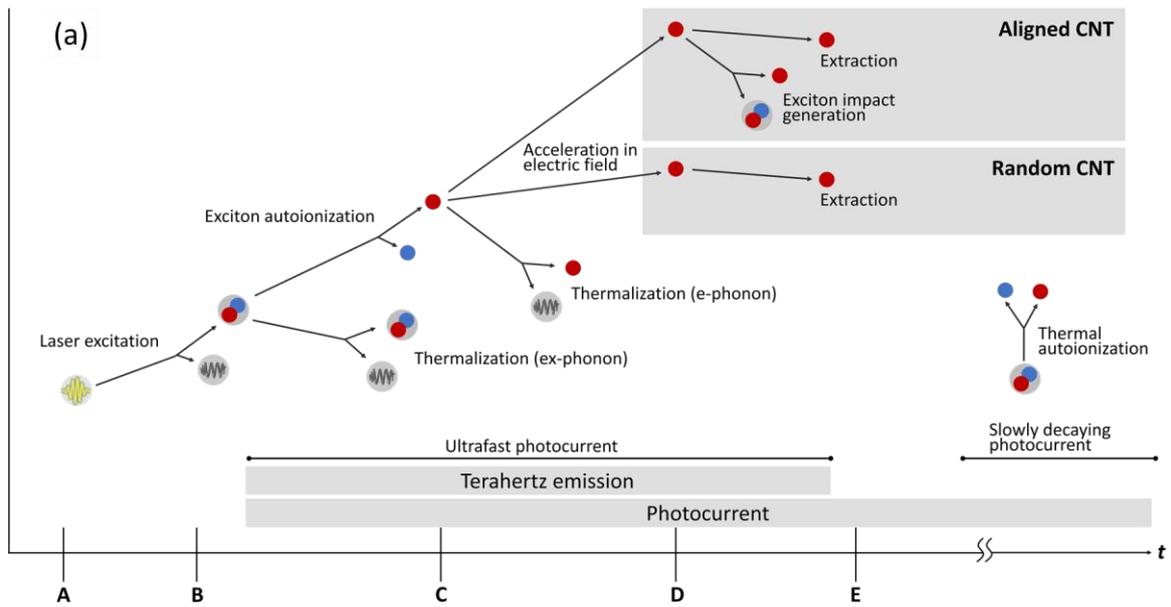

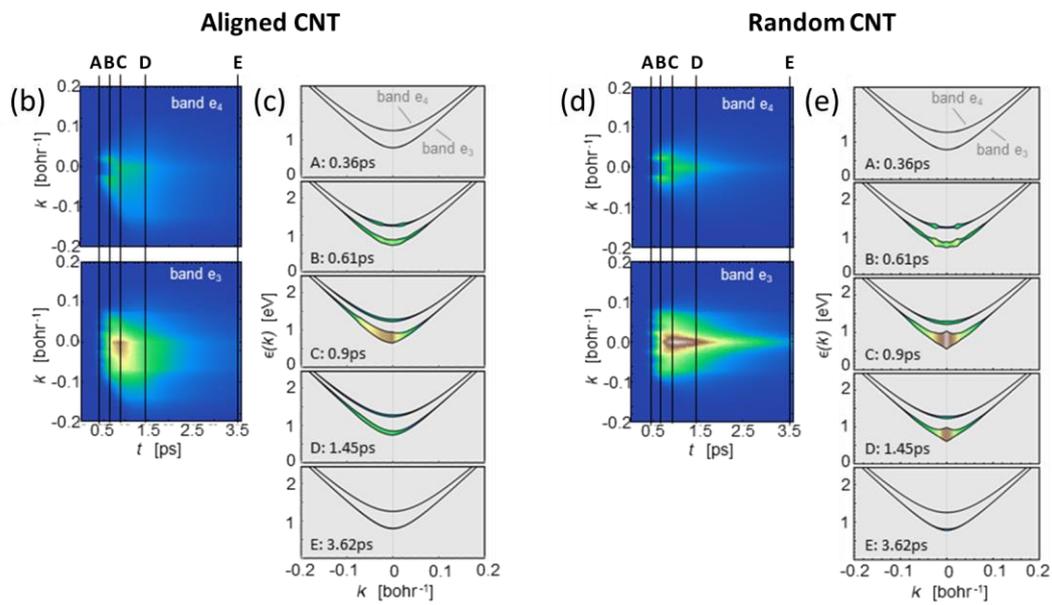

*Fig. 3.* Qualitative description of the most relevant scattering process at different time scales in aligned and random CNTs (a), with the corresponding band-resolved density plots for the electronic bands (b and d) and time snapshots of the populations (c and e), respectively.

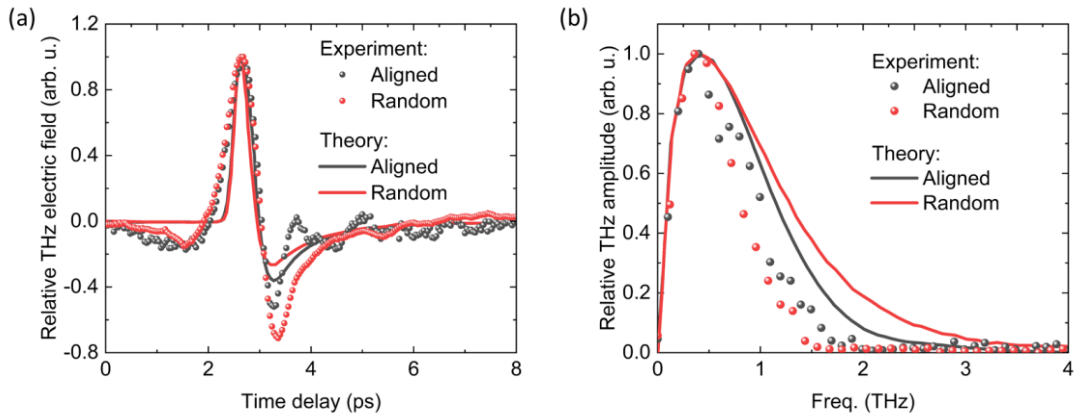

*Fig. 4.* Theoretical and experimental THz emission for aligned and random CNTs in the (a) time and (b) frequency domains.

For a closer analysis of the diffusive regime, we need to estimate how large is the asymmetric high-momentum tail of the electronic distribution at the energy threshold for exciton impact generation. To do this we compare the typical time it takes for electrons to be accelerated beyond the energy threshold (green dashed line in Figs. 5c and 5d), and the momentum-randomising total electron-phonon and electron-impurity scattering lifetime (red line). When the latter is shorter, electrons, on average, will not have time to be accelerated to the energy threshold, as they will be scattered and have their momentum randomised. This is the so-called diffusive transport regime, and lead to small high energy populations. On the other hand, when scattering lifetimes are longer, the electrons will have enough time to build up momentum under the bias and climb to higher energies. If the scattering time were infinitely long, the electrons would undergo a ballistic acceleration. In the case of aligned CNTs the scattering time is not infinity, but comparable to the time it takes the carriers to accelerate to above threshold: the motion is said to be superdiffusive, and a sizeable population will reach threshold.

Notice that yet another timescale is relevant: the momentum-resolved exciton impact ionization lifetime (blue line in Figs. 5c and 5d). If this is not sufficiently short, the vast majority of the electrons above threshold will anyway decay through electron-phonon or electron-impurity scatterings, making our probing technique inefficient. Therefore, we also compare this third timescale to understand the fraction of the electrons above threshold that is actually creating low energy excitons through impact generation. Indeed we find that in the case of random CNTs (Fig. 5c), electron-defects scattering-lifetimes are considerably shorter than the time it takes to reach the threshold energy, leading to the standard diffusive regime. The situation is importantly different in the aligned case (Fig. 5d), where the momentum-relaxing scattering-lifetimes are around half the acceleration times, a signature of a superdiffusive regime.

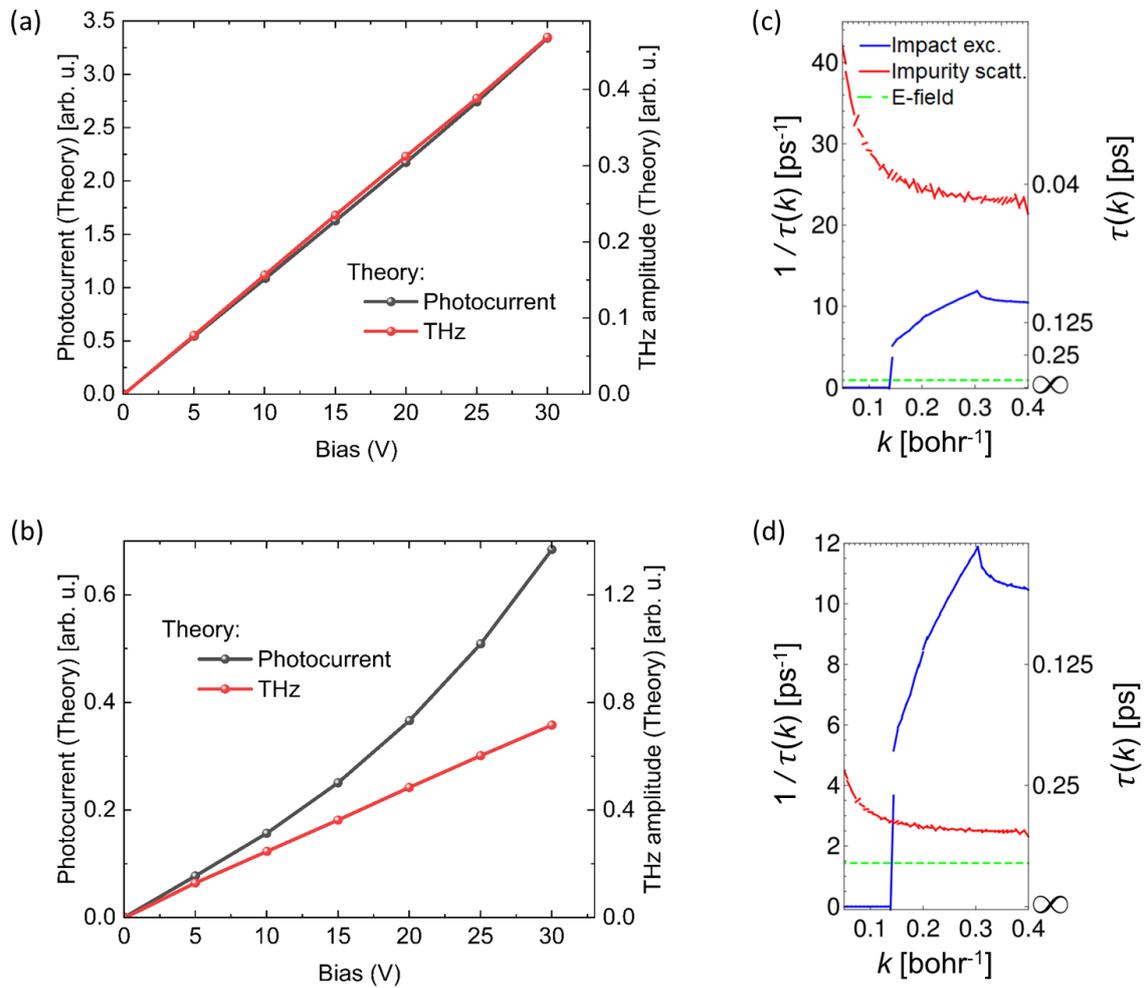

*Fig. 5.* Theoretical bias-dependence of THz amplitude and photocurrent for (a) random and (b) aligned CNTs. Calculated momentum-resolved scattering rates for electrons in the $e_3$ band for (c) random and (d) aligned CNTs. The electric field timescale is computed as the time it takes to an electron to be accelerated ballistically to the exciton impact generation threshold in the case of bias 30V.

The dynamics presented above explain well the properties of THz emission and photocurrent in our CNT-based devices. The ultrafast current contributing to THz emission linearly depends on the number of absorbed photons (Supporting Information) and the strength of the electric field (Figs. 2d and 2e). On the other hand, the photocurrent has an additional slow component produced by thermal autoionization of excitons, and the final number of these excitons is affected by bias-driven impact generation. This leads to superlinear photocurrent with bias, which we see in Figs. 2d and 2e for Samples 1 and 2 (aligned CNTs), respectively. However, for Sample 2, the deviation from linearity of the photocurrent occurs at a higher applied bias (Fig. 2e). This is due to the fact that at the same applied bias, the effective electric field across the tubes in Sample 2 is smaller compared to Sample 1 because of the mismatch between the tube orientation and the field. The superlinearity completely disappears in Sample 3 (random CNTs) because of the suppression of exciton impact generation, and instead we observe a saturation-like behaviour in both THz amplitude and photocurrent versus bias (Fig. 2f). No saturation effects were included in the modelling, which explains the discrepancy between experiment (Fig. 2f) and theory (Fig. 5a). This kind of saturation in CNTs could result from field-induced saturation effects[34], reduced mobility by optical phonon emission at high fields[35], or ionized-impurity-like scattering due to accumulated charges in CNT intersects[36], none of which were incorporated in the numerical simulations, and are beyond the scope of the present study.

We have shown how the initial electron transport in a CNT network is affected by the degree of alignment and direction relative to the applied electric field. This transport behaviour (superdiffusive or diffusive) strongly affects the kinetic energy gain of electrons by acceleration in the electric field and the production of additional excitons by impact generation, and reroutes thermalisation pathways.

These findings have great implications for the design of CNT-based optoelectronic devices. Devices that do not operate at high-frequencies, such as photodetectors, could potentially benefit from the slow thermal autoionization of impact-generated excitons for increased responsivities. For such devices, highly-aligned CNTs can be incorporated in narrow structures designed to generate high electric fields to enhance exciton impact generation while minimizing the required applied voltage. On the other hand, devices like THz emitters could benefit more from increased photon absorption to generate more charge carriers (ultrafast current) by spontaneous exciton dissociation. Without considering electron mobility and coherence effects, perhaps unaligned but thicker CNT films (which are easier to produce) are better for THz applications. Similar considerations might be viable for other low-dimensional materials where the dynamics of carriers and excitons are affected by device and material geometries.

Last but not least, we have also demonstrated the potential of THz and photocurrent measurements together with our newly developed numerical simulations in elucidating momentum-dependent ultrafast dynamics in CNTs under strong electric fields. Time-resolved pump-probe experiments are typically employed to study the ultrafast dynamics of carriers, excitons, and phonons in CNTs, but these measurements are usually limited to zero or low electric field conditions due to difficulties in probing the region with a high electric field that are usually in the micron scale. By combining THz and photocurrent experiments, we are able to access charge carrier creation at distinct time scales after photoexcitation, thereby allowing us to create an accurate microscopic model of the dynamics.

**Supporting Information**

Additional information related to the device fabrication, experimental set-up, numerical simulations and other supporting data are included in the Supporting Information. This material is available free of charge via the internet at http://pubs.acs.org.

**Acknowledgments**

N.K. and J.K. acknowledge support from the Robert A. Welch Foundation through grant number C-1509, the Air Force Office of Scientific Research through grant number FA9550-22-1-0382, the JST CREST program, Japan, through grant number JPMJCR17I5, the US National Science Foundation through grant number PIRE-2230727, and the Carbon Hub of Rice University. M.B. acknowledges Nanyang Technological University, NAP-SUG, for the funding of this research. M.W. acknowledges the Austrian Science Fund (FWF) for funding through Doctoral School W1243 Solids4Fun (Building Solids for Function) and Nanyang Technological University, NAP-SUG. K.H. acknowledges the FWF for support through project P36213. This work was partially supported by JSPS KAKENHI Grant Nos. JP18KK0140, JP23H00184, and JP22H01550, JST CREST Grant Number JPMJCR22O2, JSPS Core-to-Core Program, and Osaka University Program for Promoting International Joint Research. I.K. acknowledges support from the Iketani Science and Technology Foundation.